\documentstyle[preprint,aps,psfig]{revtex}

\def\lsim{\raise0.3ex\hbox{$\;<$\kern-0.75em\raise-1.1ex
\hbox{$\sim\;$}}}
\def\gsim{\raise0.3ex\hbox{$\;>$\kern-0.75em\raise-1.1ex
\hbox{$\sim\;$}}}
\begin{document}
\textheight = 23cm
\topmargin = -1.4cm
\baselineskip 7.6mm

\preprint{\vbox{\hbox{hep-ph/0112345}
}}
\title{
Method for Determination of $|U_{e3}|$ in Neutrino 
Oscillation Appearance Experiments
}

\author{
Takaaki Kajita$^1$\thanks{E-mail: kajita@icrr.u-tokyo.ac.jp}, 
Hisakazu Minakata$^2$\thanks{E-mail: minakata@phys.metro-u.ac.jp}
and 
Hiroshi Nunokawa$^3$\thanks{E-mail: nunokawa@ift.unesp.br}
}
\address{
$^1$\sl
Research Center for Cosmic Neutrinos,
Institute for Cosmic Ray Research, \\
University of Tokyo, Kashiwa, Chiba 277-8582, Japan}
\address{
$^2$
\sl Department of Physics, Tokyo Metropolitan University \\
1-1 Minami-Osawa, Hachioji, Tokyo 192-0397, Japan}
\address{
$^3$
\sl Instituto de F\'{\i}sica Te\'orica,
Universidade Estadual Paulista \\
Rua Pamplona 145, 01405-900 S\~ao Paulo, SP Brazil}

\maketitle
\vspace{-0.8cm}

\hfuzz=25pt
\begin{abstract}
We point out that determination of the MNS matrix element 
$|U_{e3}| = s_{13}$ in long-baseline 
$\nu_{\mu} \rightarrow \nu_e$ neutrino oscillation 
experiments suffers from large intrinsic uncertainty due 
to the unknown CP violating phase $\delta$ and sign of 
$\Delta m^2_{13}$.
We propose a new strategy for accurate determination of $\theta_{13}$;
tune the beam energy at the oscillation maximum and do the measurement 
both in neutrino and antineutrino channels. 
We show that it automatically resolves the problem of parameter 
ambiguities which involves $\delta$, $\theta_{13}$, and the sign of 
$\Delta m^2_{13}$.

\end{abstract}

\newpage

\section{Introduction}

With the accumulating evidences for neutrino oscillation in the
atmospheric\cite {SKatm}, the solar \cite {solar} and the
accelerator neutrino experiments \cite {K2K},
it is now one of the most important subjects in particle physics to
explore the full structure of neutrino masses and the lepton
flavor mixing. In particular, it is the challenging task to explore
the relatively unknown (1-3) sector of the MNS matrix \cite {MNS},
namely, $\theta_{13}$, the sign of $\Delta m^2_{13}$
and the CP violating phase $\delta$.
The only available informations to date are the upper bound on
$\theta_{13}$ from the reactor experiments \cite{CHOOZ},
and an indication for positive sign of $\Delta m^2_{13}$ by 
neutrinos from supernova 1987A \cite{MN01}.
Throughout this paper, we use the standard notation of 
the three flavor MNS matrix, in particular 
$U_{e3} = s_{13} e^{- i \delta}$, 
and define the neutrino mass-squared difference as
$\Delta m^2_{ij} \equiv m^2_{j} - m^2_{i}$.

The long baseline $\nu_{\mu} \rightarrow \nu_{e}$ 
neutrino oscillation experiment is one of 
the most promising way of measuring $\theta_{13}$. 
In particular, it is expected that the JHF-Kamioka project 
which utilizes low energy superbeam can go down to the sensitivity
$\sin^2{2 \theta_{13}} \simeq 6 \times 10^{-3}$ \cite {JHF}.
A similar sensitivity is expected for the proposed 
CERN $\rightarrow$ Frejus experiment \cite{CERNFR}.
Although a far better sensitivity is expected to be achieved in
neutrino factories \cite {nufact}, it is likely that the low energy
conventional superbeam experiments are the ones which can start
much earlier. Therefore, it is of great importance to examine 
how accurately $\theta_{13}$ can be determined in this type of 
experiments.

In this paper, we point out that determination of 
$\sin^2{2 \theta_{13}}$
by using only neutrino channel suffers from large intrinsic
uncertainty of $\pm$ (30-70) \% level due to the unknown CP
violating phase $\delta$ and the undetermined sign of 
$\Delta m^2_{13}$. It should be noted that
the intrinsic uncertainty exists on top of the usual experimental
(statistical and systematic) errors.
To overcome the problem of the intrinsic uncertainty, we suggest a
new strategy for determination of $\theta_{13}$ by doing appearance
experiments utilizing both antineutrino and neutrino beams.
Our proposal is a very simple one at least at the conceptual level;
tune the beam energy to the oscillation maximum and run the
appearance experiments in both
$\bar{\nu}_{\mu} \rightarrow \bar{\nu}_{e}$ and
$\nu_{\mu} \rightarrow \nu_{e}$ channels.

We will show that it not only solves the problem of intrinsic
uncertainty mentioned above but also resolves the
($\delta - \theta_{13}$) two-fold ambiguity discussed in
Ref. \cite {BurguetC}.
Furthermore, it does not suffer from possible ambiguity due to
the unknown sign of $\Delta m^2_{13}$, the problem first addressed
in Refs. \cite {MNjhep01,nufact01}.\footnote{
Our new strategy and these results were announced in
the 8th Tokutei-RCCN workshop \cite {TOKU-RCCN.mina}.
}
We are aware that there are combined ambiguities to be resolved
(even ignoring experimental uncertainties) to determine a complete
set of parameters including
$\delta$, $\theta_{13}$, and the sign of $\Delta m^2_{13}$,
which are as large as four-fold \cite {taup2001.mina}.
We take experimentalists' approach to the ambiguity problem
and try to resolve them one by one, rather than developing
mathematical framework for the simultaneous solutions.
The most important issue here is again to accurately determine
$\theta_{13}$, because then all the combined ambiguities
will be automatically resolved, as we will show below.

\section{Intrinsic uncertainty in determination of $\theta_{13}$
due to CP violating phase}

Let us clarify how large uncertainty is expected for determination
of $\theta_{13}$ due to our ignorance of $\delta$ in the 
$\nu_{\mu} \rightarrow \nu_{e}$ appearance experiment.  
To achieve intuitive understanding of the issue we use the 
CP trajectory diagram introduced in previous papers 
\cite {MNjhep01,nufact01}.
Plotted in Fig. 1 are the CP trajectory diagrams
in bi-probability space spanned by 
$P(\nu) \equiv P(\nu_\mu\to\nu_e) $ and 
$ P(\bar{\nu}) \equiv P(\bar{\nu}_\mu\to\bar{\nu}_e)$ 
averaged over Gaussian distribution 
(see next paragraph) 
with three values of $\theta_{13}$,
$\sin^2{2 \theta_{13}} = 0.05$ and 0.02 for 
$\Delta m^2_{23} > 0$ case and 
$\sin^2{2 \theta_{13}} = 0.064$ for 
$\Delta m^2_{23} < 0$ case.
Since we assume $|\Delta m^2_{23}| \gg |\Delta m^2_{12}|$ 
the sign of $\Delta m^2_{23}$ is identical with that of 
$\Delta m^2_{13}$.
(The fourth one with $\sin^2{2 \theta_{13}} = 0.04$ is for 
our later use.) 
The values of $\sin^2{2 \theta_{13}}$ for the second and the third 
trajectories are chosen so that the maximum (minimum) 
value of $\langle P(\nu)\rangle$ of the second (third) 
trajectory coincides with about 1.1 \%, the minimum value of 
$\langle P(\nu)\rangle$ of the first trajectory. 
The remaining mixing parameters are taken as the best fit value
of the Super-Kamiokande (SK) and the K2K experiments \cite {FLM01},
$|\Delta m^2_{23}| \equiv \Delta m^2_{atm} = 
3 \times 10^{-3}$ eV$^2$,
and the typical ones for the large mixing angle (LMA) MSW solar 
neutrino solution as given in the caption of Fig. 1.

While we focus in this paper on the JHF experiment with 
baseline length of 295 km, JAERI-Kamioka distance, 
many of the qualitative features of our results remains valid 
also for the CERN-Frejus experiment. 
Throughout this paper we take the neutrino energy distribution of 
Gaussian form with width of 20 \% of the peak energy. 
Of course, it does not represent in any quantitatively accurate 
manner the effects of realistic beam energy spread and the 
energy dependent cross sections. But we feel that it is sufficient 
to make the point of this paper clear, illuminating our 
new strategy toward accurate determination of $\theta_{13}$.

Suppose that a measurement of appearance events gives us the
value of the oscillation probability
$\langle P(\nu) \rangle \simeq 1.1$ \%. 
Then, it is obvious from Fig. 1 that a full range of values of
$\sin^2{2 \theta_{13}}$ from 0.02 to 0.064 are allowed 
(even if we ignore experimental errors) due to 
our ignorance to the CP violating phase $\delta$ and 
the sign of $\Delta m^2_{13}$.\footnote
{
It may be worth to remark the following:
Low energy neutrino oscillation experiments with superbeams
are primarily motivated as a result of the search for the place
where CP violating effects are comparatively large and easiest
to measure \cite {lowecp}. See e.g., \cite {cp-matter} for
works preceding to \cite {lowecp}.
Unfortunately, this large effect of
$\delta$ is the very origin of the above mentioned large intrinsic
uncertainty in determination of $\theta_{13}$.}
If we know that the sign is positive, for example, the uncertainty 
region would be limited to 0.02-0.05, which is still large.

Let us estimate in a syatematic way the uncertainty in the 
determination of $\theta_{13}$
due to the CP violating phase $\delta$. To do this we rely on 
perturbative formulae of the oscillation probabilities
$ P(\nu)$ and $P(\bar{\nu})$
which are valid to first order in the matter effect \cite {AKS}.
With relatively short baseline $\sim 300$ km or less the
first-order formula gives reasonably accurate results.
Ignoring O($\sin^3{2 \theta_{13}}$) terms the formula can be
written with use of the notation
$\Delta_{ij} \equiv \frac{\Delta m^2_{ij} L}{2E}$ 
($L$ and $E$ denote baseline length and neutrino energy, respectively)
in the form
\begin{equation}
P(\nu/\bar{\nu}) = P_{\pm} \sin^2{2 \theta_{13}} +
2Q \sin{2 \theta_{13}}
\cos{\left( \frac{\Delta_{13}}{2} \pm \delta \right)}
\label{prob}
\end{equation}
where
\begin{equation}
P_{\pm}(\Delta_{13}) = s^2_{23} \left[
\sin^2{\left(\frac{\Delta_{13}}{2}\right)} -
\frac {1}{2} s^2_{12} \Delta_{12}
\sin{\left(\Delta_{13}\right)}
\pm
\left( \frac{2 Ea}{\Delta m^2_{13}} \right)
\sin^2{\left(\frac{\Delta_{13}}{2}\right)}
\mp
\frac{aL}{4}
\sin{\left(\Delta_{13}\right)} \right],
\end{equation}
\begin{equation}
Q = c_{12}s_{12}c_{23}s_{23}
\Delta_{12} \sin{\left(\frac{\Delta_{13}}{2}\right)},
\end{equation}
where $a = \sqrt{2} G_F N_e$ denotes the index of refraction 
in matter with $G_F$ being the Fermi constant and $N_e$ a constant 
electron number density in the earth. 
The $\pm$ signs in $P_{\pm}$ refer to the neutrino and the
antineutrino channels, respectively.

The maximum and the minimum of $P(\nu)$ 
for given mixing paramters, neutrino energy and baseline
is obtained at
$\cos{\left(\delta + \frac{\Delta_{ij}}{2}\right)} = \pm 1$.
Then, the allowed region of $\sin{2 \theta_{13}}$ for a given
value of $P(\nu)$, assuming blindness to the sign of  
$\Delta m^2_{13}$, is given by
\begin{equation}
\frac{\sqrt{Q^2 + P_{+}(\Delta_{13})P(\nu)} - |Q|}
{P_{+}(\Delta_{13})} \leq
\sin{2 \theta_{13}} \leq
\frac{\sqrt{Q^2 + P_{+}(-\Delta_{13})P(\nu)} + |Q|}
{P_{+}(-\Delta_{13})}
\end{equation}
In Fig.~2 presented is the allowed region of $\sin^2{2 \theta_{13}}$
for a given value of measured oscillation probability $P(\nu)$. 
Figures (a)-(c) correspond respectively to neutrino energies
(a) $E=$ 500 MeV, (b) 716 MeV (oscillation maximum), and (c) 1 GeV.
One notices that the intrinsic uncertainty is large.
It strongly depends on the value of $\sin^2{2 \theta_{13}}$ and
gradually decreases as $E$ grows.
Roughly speaking, it ranges between,
$\sim$ 45 \% (30 \%) at $\sin^2{2 \theta_{13}} = 0.1$ and
$\sim$ 80 \% (70 \%) at $\sin^2{2 \theta_{13}} = 0.01$ at
$E=500$ MeV (716 MeV).
Notice that all the results shown in the plots in this paper were 
obtained by numerically solving the neutrino evolution equation assuming 
constant matter density without using the first-order formula.  

The size of the intrinsic uncertainty must be compared with the
statistical and the systematic errors which are expected in the
actual experiments. A detailed estimation of the experimental
uncertainties is performed for the JHF experiment by Obayashi
\cite {obayashi} assuming the off-axis beam (OA2) \cite {JHF}
and running of 5 years.
The results strongly depend upon $\theta_{13}$. 
We quote the case of three typical values;
$\sin^2{2\theta_{13}} = 0.1 + 0.018 - 0.014$,
$0.03 + 0.010 - 0.007$, and
$0.01 + 0.007 - 0.006$.
The errors include not only statistical but also systematic
ones. We implemented these errors in Fig.~2b which is drawn with
the similar energy as the peak energy of OA2 beam ($\sim 780$ MeV).
We should note, however, an important difference between Fig.~2 
and the plot in \cite {obayashi}; 
the abscissa of Fig.~2 is the Gaussian averaged probability, 
whereas the corresponding axis of the plot in \cite {obayashi} 
is the number of events. Therefore, we tentatively determined 
the location of errors in Fig.~2 so that the center of the error 
bars coincide with the center of the allowed band of 
$\sin^2{2\theta_{13}}$.
Keeping this difference in mind, we still feel it informative
for the readers to display the expected experimental 
uncertainties in Fig.~2b for comparison.

Therefore, the intrinsic uncertainty due to $\delta$ and 
undetermined sign of $\Delta m^2_{13}$ is
larger than the expected experimental errors
in most of the sensitivity region for $\theta_{13}$ in
the experiment.
We note that the experimental errors are dominated by the
statistical one in phase I of the JHF-SK neutrino project 
and hence it should be improved
by a factor of $\sim 10$ in two years of running in the
phase II with a megaton water Cherenkov detector \cite{JHF}.
Thus, the intrinsic uncertainty completely dominates over
the experimental ones if one stays only on the neutrino
channel.

\section{Possible way out and the relationship with
$\theta_{13} - \delta$ ambiguity}

Let us discuss possible ways out of the uncertainty problem in the
determination of $\theta_{13}$. It is tempting to think about
seeking better resolution by adding more informations.
A natural candidate for such possibilities in this line of
thought is to do additional appearance experiment
$\bar{\nu}_{\mu} \rightarrow \bar{\nu}_{e}$
using antineutrino beam. While it strengthens constraints,
it does not completely solve the uncertainty problem even if we
ignore the experimental errors.
It is due to the inherent two-fold ambiguity which exists
in simultaneous determination of $\delta$ and $\theta_{13}$
as has been pointed out by Burguet-Castell et al. \cite {BurguetC}.
While their discussion anticipates applications to neutrino
factory, the issue of the two-fold ambiguity is in fact even
more relevant to our case because of the large effect of $\delta$
as we saw in the previous section.

The existence of the two-fold ($\theta_{13} - \delta$) ambiguity
is easy to recognize by using the CP trajectory diagram.
We show in Fig.~1 by a dash-dotted curve another trajectory drawn
with $\sin^2{2 \theta_{13}} = 0.04$
which has two intersection points with the solid curve trajectory 
with $\sin^2{2 \theta_{13}} = 0.05$.
Suppose that measurements of neutrino and antineutrino oscillation
probabilities $P(\nu)$ and $P(\bar{\nu})$ had resulted into either
one of the two intersection points. Then, it is clear that we have
two solutions, for positive $\Delta m^2_{13}$, 
($\sin^2{2 \theta_{13}}, \delta$) =
(0.04, 0.65$\pi$) and  (0.05, 0.35$\pi$)
for the upper intersection point, and
($\sin^2{2 \theta_{13}}, \delta$) =
(0.04, 1.4$\pi$) and  (0.05, 1.7$\pi$)
for the lower intersection point.
Similar two-fold ($\theta_{13} - \delta$) ambiguity
also exists for negative $\Delta m^2_{13}$ which however is 
not shown in Fig.~1.
In other word, we can draw two different CP trajectories
which pass through a point determined by given values of
$P(\nu)$ and $P(\bar{\nu})$.
This is the simple pictorial understanding of the
($\theta_{13} - \delta$) two-fold ambiguity which is uncovered
and analyzed in detail in \cite {BurguetC}.
We will show in the next two sections that the ambiguity is
automatically resolved by our proposal.

\section{New strategy for determination of $\theta_{13}$}

We now present our new strategy for determination of $\theta_{13}$
which avoids the problem of the large intrinsic uncertainty.
It is intuitively obvious from the CP trajectory diagram displayed
in Fig.~1 that if one can tune the experimental parameters so that
its radial thickness (which measures the $\cos{\delta}$ term
in Eq. (\ref{prob})) vanishes then the two-fold ambiguity is completely
resolved. 
It occurs if we tune the beam energy at the oscillation maximum 
so that $\Delta_{13} = \pi$ as is clear from Eq. (\ref{prob}). 

We explain below in more detail how it occurs and then
discuss by what kind of quantity $\theta_{13}$ is determined.
In the following discussion we assume that the mixing parameters
$|\Delta m^2_{13}| \simeq |\Delta m^2_{23}| \equiv \Delta m^2_{atm}$,
$\Delta m^2_{12} \equiv \Delta m^2_{\odot}$,
$\theta_{23}$, and $\theta_{12}$ are accurately determined by the
time of the experiment.
It is not so unrealistic assumption in view of the array of
experiments ongoing (SK, SNO, K2K, KamLAND), 
on schedule (Borexino, MINOS, OPERA), or in planned (JHF).
For example, the uncertainty in measurement of $\theta_{23}$ 
is expected to be $\delta (\sin^2{2\theta_{23}}) \simeq 0.01$
in JHF phase I \cite{JHF}.

We note that the oscillation probabilities (\ref{prob})
can be written as
\begin{eqnarray}
P(\nu) &=& A \cos{\delta} + B \sin{\delta} + C_{+}
\nonumber \\
P(\bar{\nu}) &=& A \cos{\delta} - B \sin{\delta} + C_{-}
\label{general}
\end{eqnarray}
where
$A = Q \cos{\left( \frac{\Delta_{13}}{2}\right)}$,
$B = - Q \sin{\left( \frac{\Delta_{13}}{2}\right)}$, and
$C_{\pm} = P_{\pm} \sin^2{2 \theta_{13}}$ in the present approximation.
It is easy to show from this expression that CP trajectory diagram
is elliptic in the approximation that we are working \cite {MNjhep01}.
(In fact, it is the case for all the known perturbative formulae.)
Given (\ref{general}) it is simple to observe that the CP trajectory
is a straight line at the oscillation maximum, $A=0$;
the equation obeyed by the oscillation probabilities is given as
$P(\nu) + P(\bar{\nu}) = C_{+} + C_{-}$. Moreover, the first order
matter effect cancels in $C_{+} + C_{-}$, leaving the vacuum peace
of $P_{\pm}$.
Therefore, the slope of the straight-line CP trajectory is the
same as that in vacuum, and the matter effects affects only on
the maximum and the minimum points of the straight line in
$P(\nu)$ and $P(\bar{\nu})$ coordinates.
Thus, once a set of values of $P(\nu)$ and $P(\bar{\nu})$ is given
by the experiments, one can determine $C_{+} + C_{-}$ as the 
segment of the ``CP straight line'' in the diagram, and hence
$\sin^2{2 \theta_{13}}$ to which $C_{+} + C_{-}$ is proportional.
Thus, measurement of $P(\nu)$ and $P(\bar{\nu})$ at the oscillation
maximum implies determination of $\theta_{13}$ without suffering
from any uncertainties due to unknown value of $\delta$ and 
the sign of $\Delta m^2_{13}$.

In Fig.~3 we present the thinnest trajectories with
the  tuned value of the energy 
$E = 760$ MeV for $L = 295$ km (JAERI-Kamioka distance) with 
$\sin^2{2 \theta_{13}} = 0.05$ and 0.02, by taking the other 
mixing parameters given in the caption of Fig.~1.
The energy would be $E = 716$ MeV if we sit on the oscillation 
maximum. 
It arises because the contributions from higher and lower energy 
parts around the peak energy do not completely cancel because 
of the extra 1/E factor in the $\cos{\delta}$ term for 
symmetric Gaussian beam width. Thus, we need slightly 
higher energy to have the thinnest trajectory.
It should be noted, however, that the feature highly depends 
upon the specific beam shape, and will also be affected by the 
fact that the cross section has an extra approximately 
linear E dependence.

The slightly different slope of the straight-line trajectories 
of positive and negative $\Delta m^2_{13}$ 
indicates the higher order matter effect.
This effect must be (and can be) taken into account when
one try to determine $\theta_{13}$ following the method
proposed above.

\section{Comments on the relationship with
$\theta_{13} - \delta -$ sign of $\Delta m^2_{13}$ ambiguities}

We now show that the ($\theta_{13} - \delta$) ambiguity is
automatically resolved by tuning neutrino energy at the
oscillation maximum. It must be the case because two straight-line
trajectories with the same slope do not have intersection
points.
For our purpose, it suffices to work with oscillation probability
at a fixed monochromatic beam energy because averaging over
a finite width complicates the formalism and may obscure
the essence of the problem.
It can be shown \cite {BurguetC} that the difference between 
the true ($\theta_{13}$) and 
the false ($\theta_{13}'$) solutions of $\theta_{13}$ for a 
given set of $P(\nu)$ and $P(\bar{\nu})$ is given under the
small $\theta_{13}$ approximation by
\begin{equation}
\theta_{13}' - \theta_{13} = -
\frac{\sin{\delta} - z \cos{\delta}}{1 + z^2}
\frac{2Q}{P_{-} - P_{+}}
\sin{\left( \frac{\Delta_{13}}{2} \right)}
\label {difftheta_{13}}
\end{equation}
where
\begin{equation}
z = \frac{P_{-} + P_{+}}{P_{-} - P_{+}}
\tan{\left( \frac{\Delta_{13}}{2} \right)}
\end{equation}
Hence, the difference vanishes at the oscillation maximum,
$\Delta_{13} = \pi$, which means $z \rightarrow \infty$.
Thus, no ($\theta_{13} - \delta$) ambiguity exists at the
oscillation maximum as expected.

It should be emphasised that our strategy of tuning beam
energy at the oscillation maximum is not affected by the
ambiguity correlated with the sign of $\Delta m^2_{13}$
which is discussed in Ref. \cite {MNjhep01}.
It is because the matter effect split the straight-line 
CP trajectories of positive and negative $\Delta m^2_{13}$ 
toward the direction of the line itself in first order of 
the matter effect. The possible correction comes from higher
order matter effect which is small in the relatively short
baseline of the JHF (as well as the CERN $\rightarrow$ Frejus)
experiment, as shown in Fig.~3. The effect can be easily
taken care of in the actual determination of $\theta_{13}$.

\section{Concluding remarks}

In this paper, we proposed a new strategy for accurate
determination of $\theta_{13}$ without suffering from the
intrinsic ambiguity due to unknown value of $\delta$.
That is, tune the beam energy at the thinnest CP trajectory
and do the measurement both in neutrino and antineutrino channels.
We have shown that our new strategy completely resolves the
ambiguities in the determination of $\theta_{13}$ due to $\delta$
and due to the sign of $\Delta m^2_{13}$ within the
experimental accuracy attainable in such experiments.

One of the proposal which could be extracted from 
the strategy described in this paper is a possibility of
having $\bar{\nu}_{\mu}$ beam as early as possible.
It would be the promising option for the case of relatively 
large $\sin^2{2\theta_{13}}$, say, within a factor of 2-3 smaller 
than the CHOOZ bound. In this case, the 
$\nu_{\mu} \rightarrow \nu_e$ appearance events 
can be easily established in a few years of running of 
next generation neutrino oscillation experiments.
Then, the uncertainties in determination of $\theta_{13}$ would
be greatly decreased by switching to $\bar{\nu}_{\mu}$ beam rather
than just running with the $\nu_{\mu}$ beam.

What would be the implication of our strategy to the determination of
$\delta$? The tuning of beam energy at thinnest trajectory in fact 
also provides a good way of measuring $\delta$.\footnote
{
Tuning of beam energy at the oscillation maximum itself has been
proposed before for differing reasons from ours. First of all, 
it is preferred experimentally because it maximizes 
disappearance of $\nu_{\mu}$ as well as the number of electron 
appearance events \cite{JHF}. 
The tuning of beam energy to the oscillation 
maximum for measurement of CP violating phase $\delta$ was  
proposed by Konaka for the purpose of having maximal CP-odd
($\sin{\delta}$) term at the energy \cite{konaka,JHF}.
}
The ambiguity ($\delta \rightarrow \pi - \delta$), however, is 
unresolved and it would necessitate supplementary measurement 
either by using ``fattest" trajectory configuration \cite{MNjhep01}, 
or by second detector with different baseline distance 
\cite {BurguetC}.
We should emphasize that once $\theta_{13}$ is measured 
accurately there is no more intrinsic ambiguities in 
determination of $\delta$. 
We have explicitly shown that ($\delta - \theta_{13}$) 
ambiguity is resolved. 
The only ambiguity which would survive 
(from the viewpoint of determination of $\delta$)
would be the accidental one that arises in a correlated way 
($\delta$ $-$ sign of $\Delta m^2_{13}$), which is nothing 
but the remnant of ($\delta \rightarrow \pi - \delta$)
degeneracy in vacuum \cite {MNjhep01}. But it is also 
resolved by either one of the two second measurements 
mentioned above.

\vskip0.5cm

\noindent
Nore added:

While this paper was being written, we became aware of the
paper by Barger et al. \cite{BMW01} whose results partially
overlaps with ours. However, most of the ambiguities
discussed in the paper will be gone once $\theta_{13}$ is
determined accurately, as we noted above.

\acknowledgments

We thank  Takashi Kobayashi and Yoshihisa Obayashi
for valuable informative correspondences on low energy
neutrino beams, detector backgrounds, and $\theta_{13}$
sensitivity in the JHF experiment.
This work was supported by the Brazilian funding agency 
Funda\c{c}\~ao de Amparo \`a Pesquisa do Estado de 
S\~ao Paulo (FAPESP), 
and by the Grant-in-Aid for Scientific Research
in Priority Areas No. 12047222, Japan Ministry
of Education, Culture, Sports, Science, and Technology.


\vglue 1.0cm 
\begin{figure}[ht]
\centerline{\protect\hbox{
\psfig{file=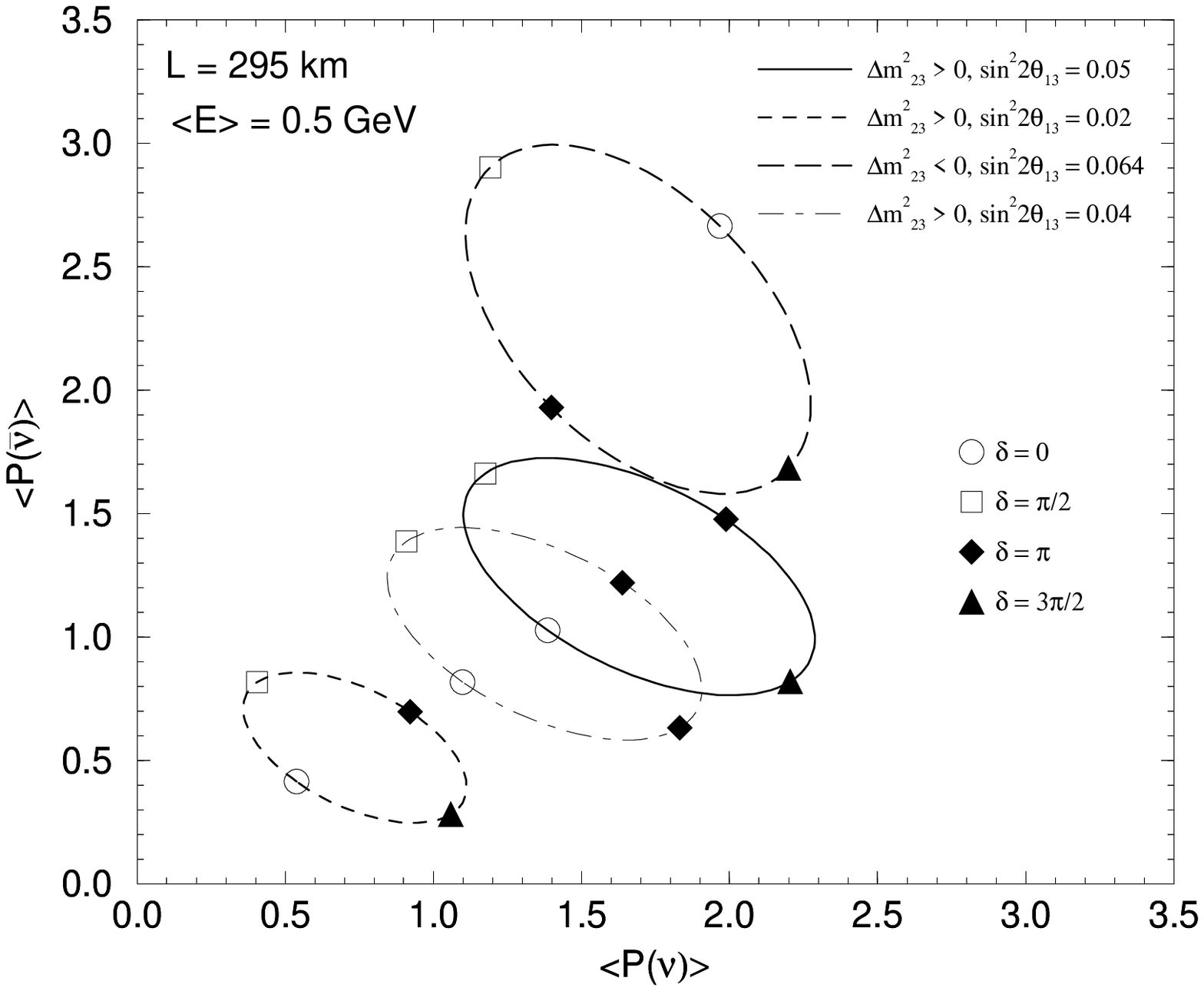,height=13cm,width=14.cm}
}}
\vglue 0.5cm 
\caption
{CP trajectory diagrams showing the contours of 
$\langle P(\nu)\rangle \equiv \langle P(\nu_\mu\to\nu_e) \rangle$ and 
$\langle P(\bar{\nu})\rangle 
\equiv \langle P(\bar{\nu}_\mu\to\bar{\nu}_e) \rangle$ 
as a function of $\delta$. 
The Gaussian energy distribution of neutrino beam with 
$\langle E \rangle = $ 0.5 GeV with width $\sigma$ = 0.1 GeV 
is assumed and the baseline length is taken as $L$ = 295 km.
The mixing parameters are fixed to be 
$\Delta m^2_{23} = \pm 3 \times 10^{-3}$ eV$^2$, 
$\sin^22\theta_{23} = 1.0$, 
$\Delta m^2_{12} = 6.2\times 10^{-5}$ eV$^2$, 
$\tan^2\theta_{12} = 0.35$.  
We take the matter density as $\rho = 2.8$ g/cm$^3$ 
and the electron fraction as $Y_e$ = 0.5. 
}
\end{figure}

\newpage 

\begin{figure}[ht]
\centerline{\protect\hbox{
\psfig{file=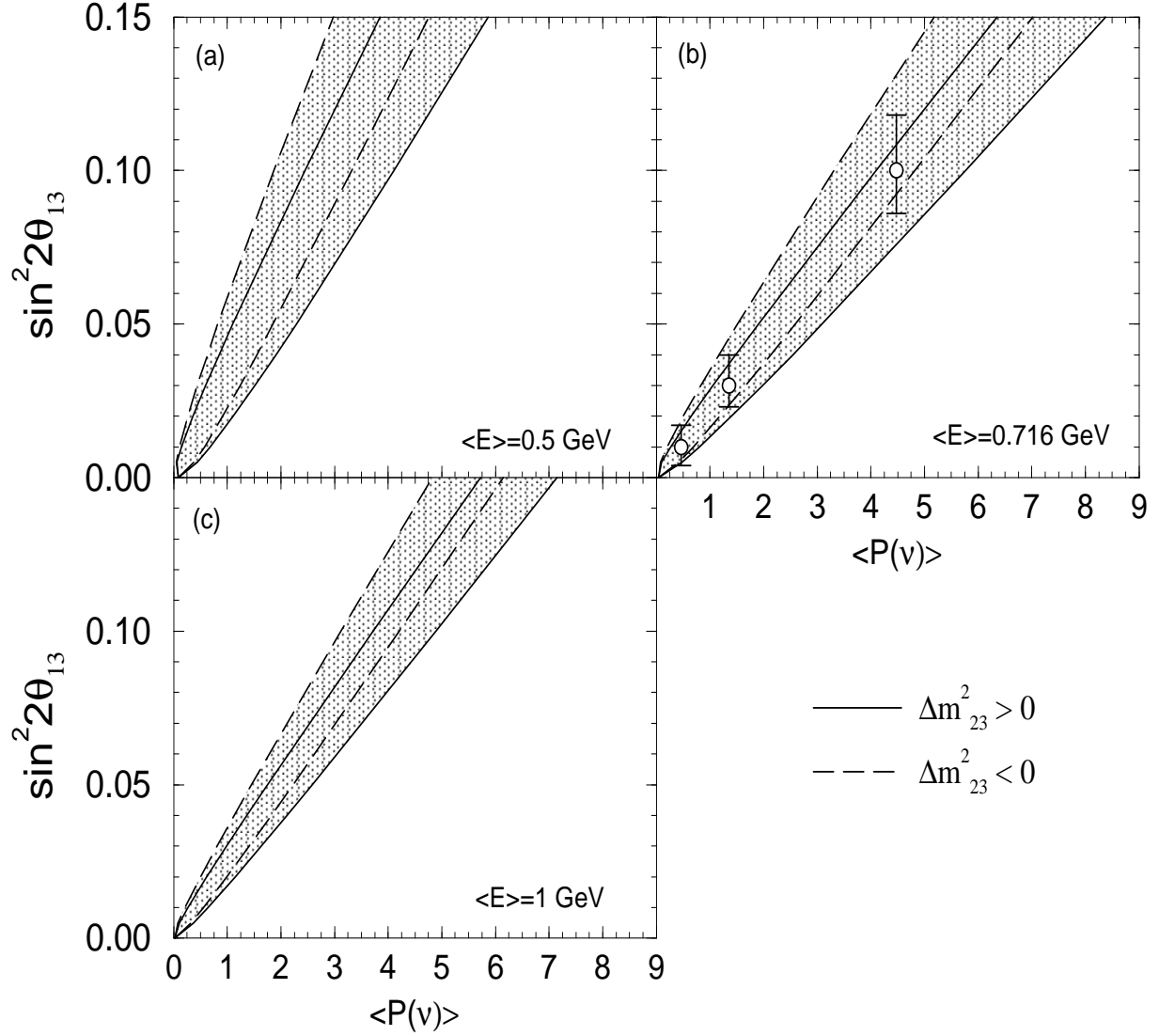,height=15cm,width=16.cm}
}}
\vglue 0.3cm 
\caption
{
Allowed region of $ \sin^22\theta_{13}$ is shown as a shaded 
strip for given values of $\langle P(\nu_\mu \to \nu_e)\rangle$ 
(given in \%) 
assuming Gaussian energy distribution of neutrino beam 
centered at 
$\langle E \rangle = $ (a) 0.5, (b) 0.716, and (c) 1.0 GeV 
with 20 \% width $\sigma$ of $\langle E \rangle $
for L = 295 km. 
If the sign of $\Delta m^2_{23}$ is known, the allowed region 
is within the solid ($\Delta m^2_{23}>0$) and the dashed 
($\Delta m^2_{23}<0$) lines. 
The other mixing parameters and the matter density are taken 
as in Fig.~1.
}
\end{figure}

\newpage 

\begin{figure}[ht]
\centerline{\protect\hbox{
\psfig{file=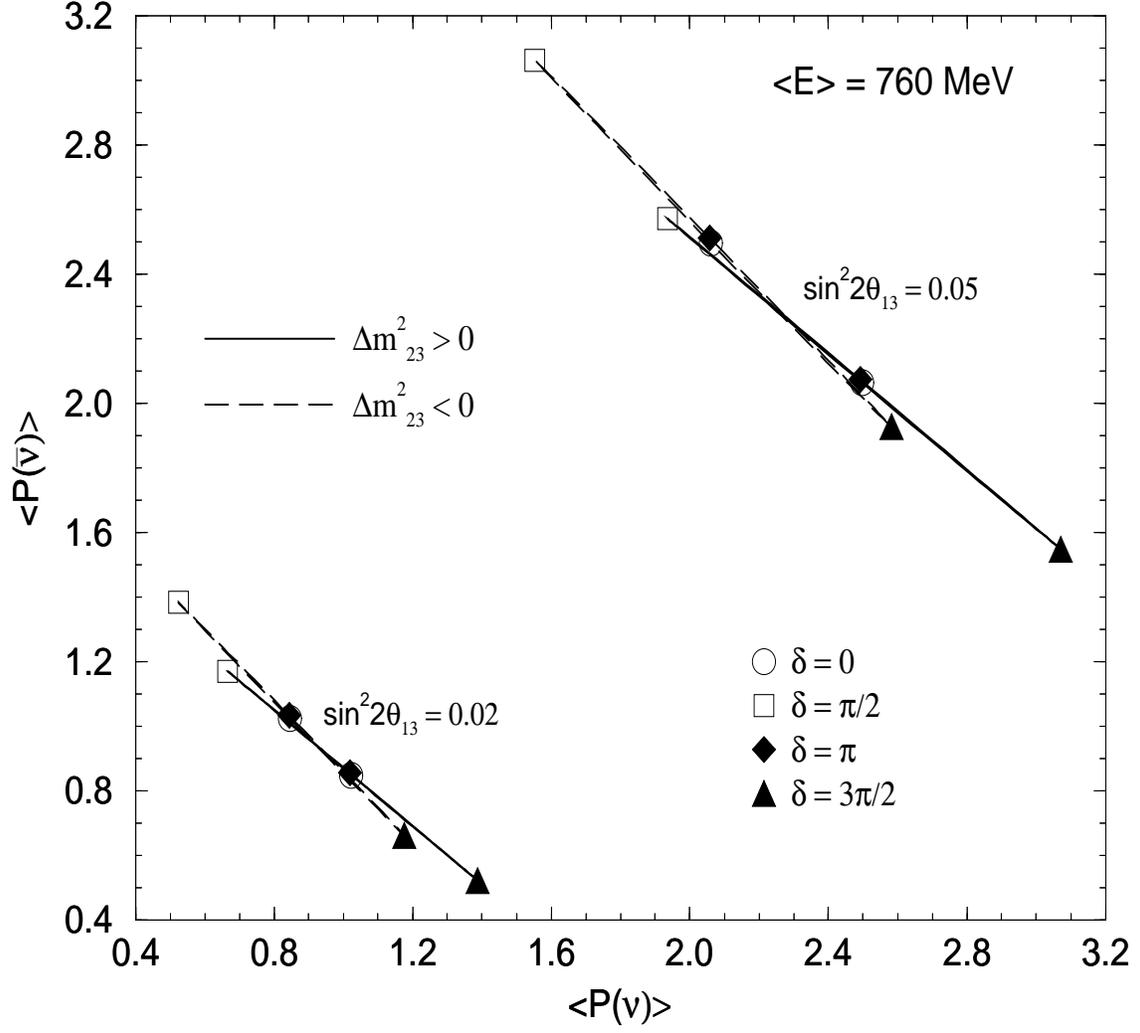,height=14cm,width=15.cm}
}}
\vglue 0.5cm 
\caption
{
The thinnest CP trajectories for a tuned peak energy 
for $\sin^2{2\theta_{13}} = 0.05$ and 0.02. 
The beam profile, the mixing parameters and the matter density 
are taken as in Fig.~1.
}

\end{figure}

\end{document}